\documentclass[final,3p,times, 11pt]{elsarticle}
\usepackage{graphicx, color}
\usepackage{bm}

\usepackage{amsmath}

\journal{Aerospace Science and Technology}

\begin{document}

\begin{frontmatter}

\title{Variational mode decomposition analysis of the relationship between low-frequency shock-wave oscillations and buffet cells}

\author[label1]{Yuya Ohmichi\corref{mycorrespondingauthor}}
\address[label1]{Aviation Technology Directorate, Japan Aerospace Exploration Agency, 7-44-1 Jindaiji-Higashi, Chofu, Tokyo 182-8522, Japan}
\cortext[mycorrespondingauthor]{Corresponding author}

\ead{ohmichi.yuya@jaxa.jp}

\author[label1]{Yosuke Sugioka}

\begin{abstract}
This study investigates the relationship between low-frequency shock-wave oscillations and buffet cells on 
the main wing of the NASA common research model.
The flow conditions were set at a Mach number of 0.85, 
Reynolds number of 2.27 $\times$ 10$^6$, and angle of attack of 4.82$^{\circ}$.
Buffet cells are cellular patterns with spanwise periodicity that propagate toward the wing tip, with nondimensional frequencies (Strouhal numbers) of 0.2--0.6, higher than those of shock oscillations associated with transonic buffet.
However, the physical mechanisms driving buffet cells and their interaction with low-frequency shock motion are not fully understood.
This study employs variational mode decomposition-based nonstationary coherent structure (VMD-NCS) analysis applied to unsteady pressure-sensitive paint measurement data.
The results reveal that the amplitude of buffet cells is not directly coupled to the low-frequency shock-wave oscillations but instead shows correlation with the boundary layer separation state.
Strong buffet cell fluctuations persist in regions of sustained separation (midspan), while their amplitude varies with local separation/reattachment dynamics in regions of intermittent separation (outboard). These observations suggest that shock-waves are important to generate the adverse pressure gradient necessary for boundary layer separation, but their influence on buffet cells depends on the resulting separation behavior. This insight from experimental observations in this study is consistent with previous computational studies suggesting that buffet cell behavior depends on boundary layer separation characteristics. Moreover, this study demonstrates the effectiveness of VMD-NCS analysis for investigating nonstationary fluid phenomena with multiple spatial and temporal patterns.
\end{abstract}

\end{frontmatter}

\section{Introduction}
Transonic buffet is a phenomenon in which shock-waves on aircraft wing surfaces oscillate due to interaction with the boundary layer in the transonic regime, causing significant fluctuations in aerodynamic forces that can lead to structural vibrations.
Because the transonic buffet phenomenon imposes constraints on the flight envelope, it remains an active area of research aimed at understanding its mechanisms and related flow behavior.

The well known physical model for the mechanism of the transonic buffet phenomenon is Lee's model \cite{Lee1990}.
This model assumes a feedback mechanism whereby disturbances generated at the shock foot travel downstream through the boundary layer to the trailing edge, resulting in the formation of Kutta waves that propagate upstream and interact with the shock-wave.
Other studies using global stability analysis \cite{Crouch2007} or resolvent analysis \cite{Iwatani2023} have shown that transonic buffet emerges as a linear instability mode, and they have revealed the phenomenon of pressure waves generated at the shock foot that propagate along the shock-wave.
Furthermore, recent studies have investigated buffet mechanisms in laminar conditions \cite{Dandois2018, Kojima2020, Moise2022, Moise2024}.
In particular, Moise et al. \cite{Moise2024} investigated the link between transonic buffet and low frequency oscillation phenomenon that occurs in the incompressible regime, thereby questioning the necessity of shock-wave existence for buffet occurrence.
As described above, the mechanisms of transonic buffet remain under debate.
Nevertheless, studies aimed at mitigating buffet \cite{Gao2016, Chen2022} and accurate and quick prediction of buffet phenomenon \cite{Petrocchi2024,Ma2024} have also progressed.
For comprehensive overviews of transonic buffet phenomena, the reader is referred to the reviews by Lee \cite{Lee2001} and Giannelis et al. \cite{Giannelis2017}.

While these studies have advanced the understanding and control of the two-dimensional buffet phenomenon, the transition to swept-wing configurations introduces additional complexities arising from the three-dimensional nature of the flow.

For swept wings, recent studies \cite{Iovnovich2015, Koike2016, Dandois2016} have reported the occurrence of cellular patterns with spanwise periodicity, known as buffet cells, that propagate towards the wing tip. While two-dimenstional buffet exhibits a low-frequency periodic shock oscillation at Strouhal numbers ($St = f c_{\rm MAC}/U_\infty$, where $f$, $c_{\rm MAC}$, and $U_\infty$ represent the frequency, the mean aerodynamic chord length, and freestream velocity, respectivly) of $St = 0.05$ to $0.1$, buffet cells are characterized by a broadband frequency range of $St = 0.2$ to $0.6$, as first observed by Roos \cite{Roos1985}.

Global stability analyses of flow fields around the NASA common research model (NASA-CRM) \cite{Timme2020,Sansica2023} have demonstrated that the buffet cell phenomenon manifests as a global stability mode.
Plante et al. \cite{Plante2021} highlighted the similarity between buffet cells and stall cells, suggesting through global stability analysis that they share the same origin, thereby indicating that shock-waves are not the direct mechanism driving buffet cells.
However, accurately reproducing experimental results through numerical simulations continues to be challenging, with ongoing attempts utilizing large eddy simulation (LES) \cite{Lutz2023, Tamaki2024,Lusher2024}.

Unsteady pressure-sensitive paint (PSP) \cite{Liu2021} measurement is an optical technique that provides high temporal and spatial resolution pressure data. It has been applied to flow fields from low \cite{Nakakita2011,Ohmichi2022} to hypersonic \cite{Running2019} speeds and is effective for analyzing buffet and buffet cell phenomena \cite{Koike2016,Masini2020,Sugioka2021}.
Koike et al. \cite{Koike2016} and Sugioka et al. \cite{Sugioka2021} measured unsteady pressure fluctuation distributions on the main wing of the NASA-CRM.
Masini et al. \cite{Masini2020} applied PSP to the RBC12 half-model.
All these results confirm the occurrence of cellular structures propagating outboard with frequencies of $St=0.2$ to $0.6$. 

Nevertheless, experimental investigations into the mechanism of buffet cells remain limited. The analysis of buffet cells with broadband frequency characteristics is challenging due to their nonstationary nature.
Therefore, although computational studies \cite{Plante2021, Lusher2024} have offered insights into potential mechanisms, experimental investigation into the relationship between low-frequency shock motion and buffet cells is still lacking.

For analyzing unsteady fluid phenomena, modal analysis methods such as proper orthogonal decomposition (POD) \cite{Lumley1967}, dynamic mode decomposition (DMD) \cite{Schmid2010} and spectral POD (SPOD) \cite{Towne2018} are commonly employed \cite{Taira2017}.
Ohmichi et al. \cite{Ohmichi2018, Ohmichi2022ttls} and Masini et al. \cite{Masini2020} successfully extracted DMD modes representing low-frequency shock-wave oscillations and buffet cell structures. 
Lusher et al. \cite{Lusher2024} applied SPOD to LES data of flow around an unswept wing, reporting that modes representing cellular patterns were extracted even in cases without sweep angle when the mean flow separation was significant.
However, these modal analysis methods have inherent limitations: DMD and SPOD assume flow periodicity, restricting their application to non-periodic phenomena such as buffet cells, while conventional POD suffers from frequency mixing problems that complicate physical interpretation \cite{Noack2016}.

This study employs variational mode decomposition (VMD) \cite{Dragomiretskiy2014a, Rehman2019}.
VMD was developed as a signal processing method for nonstationary signals.
Recently, applications to fluid data have been proposed, including Reduced-order VMD \cite{Liao2023} and VMD-based nonstationary coherent structure (VMD-NCS) analysis \cite{Ohmichi2024}.
Unlike the aforementioned modal analysis methods, VMD-based methods address their limitations: they do not require the periodicity assumption of DMD/SPOD, reduce the frequency mixing issues of POD through band-limited decomposition, and allow each mode to exhibit time-varying amplitude and frequency.
While these capabilities are promising, their applications remain limited.

The objective of this study is to elucidate the relationship between low‑frequency shock oscillations and buffet cells on a swept wing by applying the VMD-NCS approach to experimental PSP data obtained by Sugioka et al. \cite{Sugioka2021}. Using VMD-NCS, we extract spatiotemporal patterns associated with both low‑frequency shock‑wave oscillations and buffet cells. We then examine how the shock-wave motion at low frequencies relates to the instantaneous amplitude of buffet cells.
Through this analysis, this study provides experimental evidence supporting recent computational studies \cite{Plante2021, Lusher2024}, demonstrating that the amplitude of buffet cells is not directly coupled to shock motion but is more closely correlated with the boundary layer separation state. Furthermore, this study highlights the effectiveness of the VMD-NCS analysis in investigating nonstationary fluid phenomena.

\section{Methodology}
\subsection{Hilbert transform}
The Hilbert transform is employed in this study for two primary purposes: to analyze the instantaneous amplitude and phase characteristics of nonstationary signals, and  in the formulation of the objective function for VMD.
In signal processing, the Hilbert transform provides a mathematical framework for constructing the analytic signal, which enables the separation of amplitude and phase information from oscillatory signals.

The Hilbert transform $H[f](t)$ of a real signal $f(t)$ is defined as:
\begin{equation}
H[f](t) = \frac{j}{\pi t} * f(t) = \frac{1}{\pi} \text{P.V.}\int_{-\infty}^{\infty} \frac{f(\tau)}{t-\tau} d\tau \label{eq:Hft}
\end{equation} 
where $j$ is the imaginary unit ($j = \sqrt{-1}$), $\ast$ denotes convolution, and P.V. represents the Cauchy principal value.
The analytic signal $z(t)$ is constructed as:
\begin{equation}
z(t) = f(t) + jH[f](t)  \label{eq:zt}
\end{equation}
The instantaneous amplitude $A(t)$ and phase $\theta(t)$ are given by:
\begin{equation}
A(t) =  \sqrt{f(t)^2 + H[f](t)^2}~~~~{\rm and}~~~~\theta(t) =  \arctan\left(\frac{H[f](t)}{f(t)}\right) \label{Eq:amp_phase}
\end{equation}
The instantaneous amplitude represents the temporal intensity variation of the signal and is commonly referred to as the amplitude envelope, while the instantaneous phase indicates the temporal change in phase, which characterizes the frequency modulation of the signal. The instantaneous frequency can be further derived as the time derivative of the instantaneous phase.

The implementation of the Hilbert transform used in this study was based on the SciPy library \cite{2020SciPy-NMeth}. 
No preprocessing of the input signal, including windowing or detrending, was applied prior to the Hilbert transform.

\subsection{Multivariate variational mode decomposition}
Multivariate variational mode decomposition (MVMD) \citep{Rehman2019} is an extension of the variational mode decomposition (VMD) algorithm \citep{Dragomiretskiy2014a} for handling multivariate signals. The method decomposes a multivariate signal $\boldsymbol{f}(t) = [f_1(t), f_2(t), ..., f_C(t)]^T$ into $K$ band-limited MVMD modes $\boldsymbol{u}_k(t) = [u_{k,1}(t), u_{k,2}(t), ..., u_{k,C}(t)]^T$, where each MVMD mode has a center angular frequency $\omega_k$. The key idea of MVMD is to find a set of MVMD modes that minimize the sum of their frequency bandwidths while reproducing the input signal. This is achieved by solving the following minimization problem:
\begin{gather}
\min_{\{\boldsymbol{u}_k\},\{\omega_k\}} \sum_{k=1}^K \left\|\partial_t\left[\left(\delta(t) + \frac{j}{\pi t}\right) * \boldsymbol{u}_k(t)\right]e^{-j\omega_k t}\right\|_2^2 \\
 {\rm subject~to} \sum_{k=1}^K \boldsymbol{u}_k(t) = \boldsymbol{f}(t)
\end{gather}
where $\partial_t$ denotes the time derivative, and $\delta(t)$ is the Dirac delta function.
The term $\left(\delta(t) + j/(\pi t)\right) * \boldsymbol{u}_k(t)$ 
 corresponds to the analytic signal of the $k$-th mode as given by Eqs. (\ref{eq:Hft}), (\ref{eq:zt}).

The optimization problem is solved utilizing the augmented Lagrangian and alternating direction method of multipliers (ADMM) approach in the frequency-domain, which leads to the following update scheme \citep{Rehman2019}:
\begin{equation}
\hat{\boldsymbol{u}}_k^{n+1}(\omega) = \frac{\hat{\boldsymbol{f}}(\omega) - \sum_{i\neq k} \hat{\boldsymbol{u}}_i^{n}(\omega) + \hat{\boldsymbol{\lambda}}^{n}(\omega)/2}{1 + 2\alpha(\omega-\omega_k^n)^2}
\end{equation}
\begin{equation}
\omega_k^{n+1} = \frac{\int_0^\infty \omega\|\hat{\boldsymbol{u}}_k^{n+1}(\omega)\|_2^2 d\omega}{\int_0^\infty \|\hat{\boldsymbol{u}}_k^{n+1}(\omega)\|_2^2 d\omega}
\end{equation}
\begin{equation}
\hat{\boldsymbol{\lambda}}^{n+1}(\omega) = \hat{\boldsymbol{\lambda}}^n(\omega) + \tau\left(\hat{\boldsymbol{f}}(\omega) - \sum_{k=1}^K \hat{\boldsymbol{u}}_k^{n+1}(\omega)\right)
\end{equation}
where $\hat{~}$ denotes the Fourier transform and $n$ is the iteration counter.
The parameter $\alpha$ is a user-defined parameter that relates to the frequency bandwidth of MVMD modes, and $\tau$ is the parameter controlling the step size to update the Lagrangian multiplier $\hat{\boldsymbol{\lambda}}$.
This iterative process continues until convergence.
The convergence criterion was set to:
\begin{equation}
     \frac{\sum_k  \left|\hat{\boldsymbol{u}}_{k}^{n+1}(\omega) - \hat{\boldsymbol{u}}_{k}^{n}(\omega) \right|^2}{\sum_k \left| \hat{\boldsymbol{u}}_{k}^{n}(\omega) \right|^2} < \varepsilon, ~~~~{\rm for~all~} \omega \geq 0.
\end{equation}
where $\varepsilon = 10^{-6}$ in this study.
Finally, a set of MVMD modes $\boldsymbol{u}_k(t)$ is obtained as the inverse Fourier transform of $\hat{\boldsymbol{u}}_k(\omega)$, which captures distinct band-limited frequency components of the multivariate signal.

In this study, besides the convergence criterion above, we confirmed that the reconstruction error, computed by the following Eq.~\eqref{eq:reconst_error} was less than $10^{-2}$:
\begin{equation}
  E =
  \frac{
    \left\lVert \boldsymbol{f}(t) - \sum_{k} \boldsymbol{u}_{k}(t) \right\rVert_{2}
  }{
    \left\lVert \boldsymbol{f}(t) \right\rVert_{2}
  }
  \label{eq:reconst_error}
\end{equation}

The practical MVMD implementation used in this study was based on Rehman et al. \cite{Rehman2019}.
The input signal was first padded on both sides with segments equal to half its length, each segment being a mirror reflection of the original data across the corresponding boundary to suppress the boundary artifacts inherent in the discrete Fourier transform \cite{Dragomiretskiy2014a,Liao2023}, and no additional window function was applied.
In the ADMM iterations, we set $\tau = 3.0$ and confirmed that the resulting modes are insensitive to the specific value of $\tau$, provided the iterations converge.
The selection of other parameters is discussed in Sec.~\ref{sec:params}.

\subsection{Variational mode decomposition–based nonstationary coherent structure analysis}
\subsubsection{Input data}
The analysis is performed on fluctuating components obtained by subtracting the time-averaged field from instantaneous values:
\begin{equation}
q'(\boldsymbol{x},t) = q(\boldsymbol{x},t) - \bar{q}(\boldsymbol{x})
\end{equation}
where $q'$, $q$, and $\bar{q}$ represent the fluctuating component, instantaneous value, and time-averaged component, respectively. Using these fluctuating components, we construct a data matrix as:
\begin{equation}
Q' = [q'(t_1), q'(t_2), \cdots, q'(t_m)]
\end{equation}
Here, each column vector $q'(t_j)$ represents the fluctuating component of the flow field at time $t_j$. The resulting matrix $Q'$ serves as the input for the VMD-NCS analysis, where the dimensions $d \times m$ correspond to the number of spatial points ($d$) and snapshots ($m$), respectively.

\subsubsection{Dimension reduction}
Since it is computationally challenging to apply MVMD directly to high-dimensional data such as from LES or unsteady PSP measurements, we first perform dimension reduction utilizing proper orthogonal decomposition (POD):
\begin{equation}
\tilde{Q} = P^T Q'
\end{equation}
where $P$ is the POD basis matrix and $\tilde{Q}$ represents the low-dimensional representation of the data $Q'$.
We note that the conventional POD algorithm may become computationally expensive for extremely large datasets $Q'$. In such cases, online \cite{Ross2008, Ohmichi2017c} or parallelized \cite{Asada2025} algorithms  can be employed as an efficient alternative.
The subsequent analyses are performed on this reduced-dimensional data $\tilde{Q} = [\tilde{q}(t_1), \tilde{q}(t_2), \cdots, \tilde{q}(t_m)]$. Through this dimension reduction, the flow field fluctuations can be approximated as:
\begin{equation}
q'(\boldsymbol{x},t) = \sum_{i=1}^r \tilde{q}_i(t)\phi_i(\boldsymbol{x})
\end{equation}
where $r$ is the truncation rank, $\phi_i$ represents the $i$-th POD mode, and $\tilde{q}_i$ denotes its corresponding temporal coefficient. Consequently, the temporal evolution of the flow field is captured by the temporal coefficients $\tilde{q}_i(t)$.

\subsubsection{Intrinsic coherent structure}
\begin{figure}
\centerline{\includegraphics[width=0.45\textwidth]{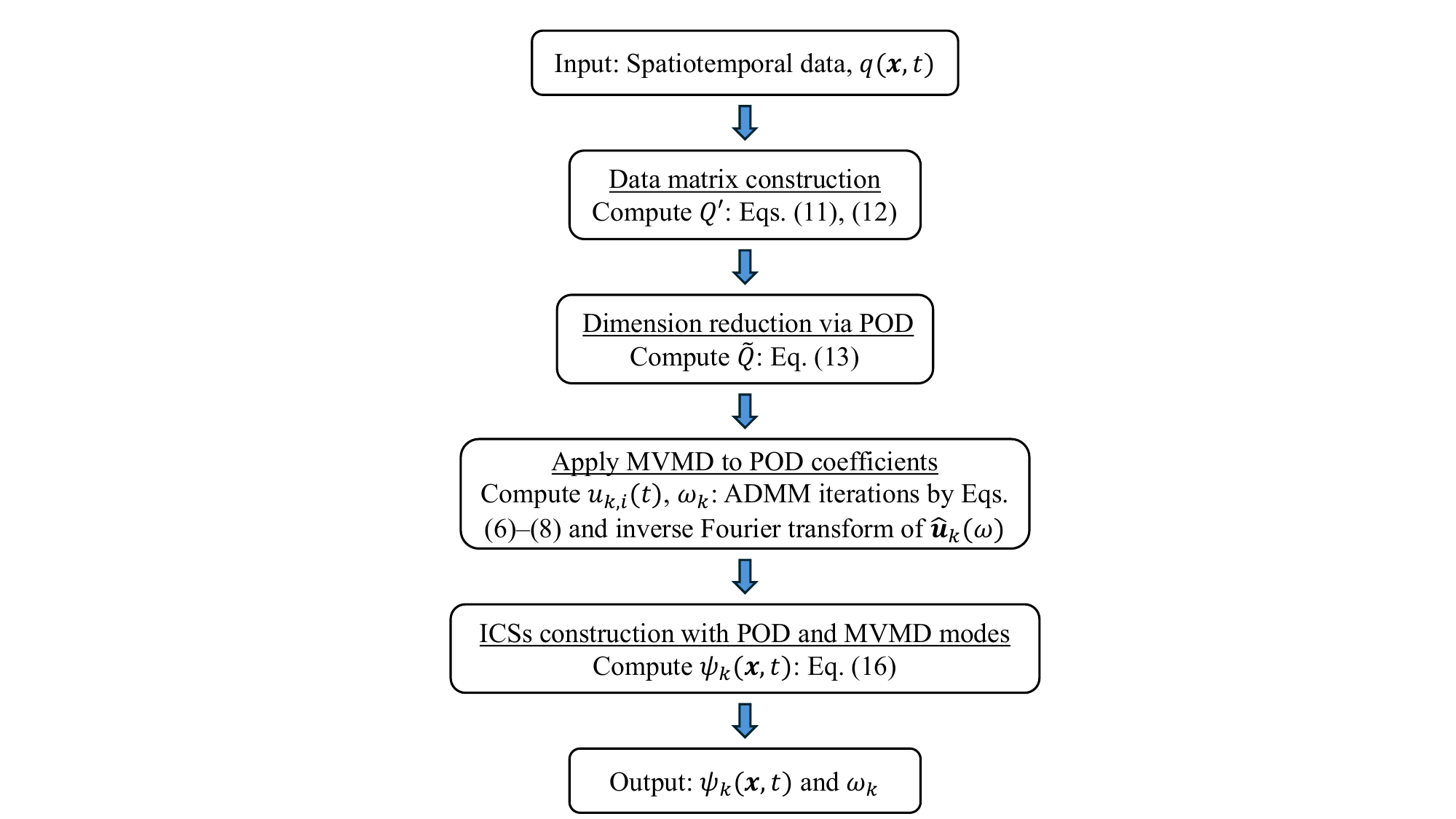}}
\caption{Flowchart of the VMD-NCS analysis pipeline for extracting ICSs from spatiotemporal data.}\label{fig:flowchart}
\end{figure}

The VMD-NCS analysis extracts spatiotemporal patterns as intrinsic coherent structures (ICSs). 
These ICSs are computed through the decomposition of the POD coefficients $\tilde{q}_i(t)$ utilizing MVMD. The temporal coefficient for the $i$-th POD mode, $\tilde q_i(t)$, is decomposed into $u_{k,i}(t)$ as:
\begin{equation}
\tilde q_i(t) =  \sum_{k=1}^{K} u_{k,i}(t) \label{eq:MVMDofPOD}
\end{equation}
Here, MVMD decomposes the POD temporal coefficients into distinct temporal patterns, each associated with a center angular frequency $\omega_k$. The $k$-th ICS $\psi_k(\boldsymbol{x},t)$ is then defined as the superposition of POD modes with the $k$-th MVMD modes as temporal coefficients:
\begin{equation}
\psi_k(\boldsymbol{x},t) = \sum_{i=1}^r u_{k,i}(t)\phi_i(\boldsymbol{x}) \label{eq:ICS}
\end{equation}
Note that the ICSs are not normalized, which differs from the treatment of POD or DMD modes.
The instantaneous flow field $q'(\boldsymbol{x},t)$ can be expressed as the superposition of these ICSs:
\begin{equation}
q'(\boldsymbol{x},t) = \sum_{k=1}^K \psi_k(\boldsymbol{x},t)
\end{equation}
We illustrate the data processing pipeline for VMD-NCS analysis in Fig. \ref{fig:flowchart}.

A key feature of ICSs is their ability to represent spatiotemporally varying structures as single ICSs, each associated with a specific center frequency while allowing for nonlinear amplitude change and temporal frequency modulation. This makes them suitable for analysing nonstationary phenomena in fluid flows.
For further details of the VMD-NCS analysis, please refer to \cite{Ohmichi2024}.

\subsubsection{Setting parameters} \label{sec:params}
The VMD-NCS analysis involves three key parameters: the truncation rank $r$, the bandwidth parameter $\alpha$, and the number of ICSs $K$. The truncation rank $r$ is determined based on the contribution rate, and in this study, is set to $r = 320$, corresponding to a 90\% contribution rate.
A previous study \citep{Ohmichi2024} revealed that large values of $\alpha$ or high $K$ settings lead to narrow-band decomposition, where the extracted ICSs effectively capture periodic phenomena but may miss nonperiodic variations. Conversely, small values of $\alpha$ or $K$ allow for broader frequency content, although extremely small values may compromise the temporal coherence of the ICSs.
Additionally, excessively large values of $\alpha$ can result in poor convergence in the ADMM iterations.
In the present study, we employ $K = 3$ and $\alpha = 1$ based on parameter studies (see Appendix A), which demonstrate that this combination provides optimal separation between low-frequency shock oscillations and buffet cells.

\section{Test case and input data}
\begin{figure}
\centerline{\includegraphics[width=0.5\textwidth]{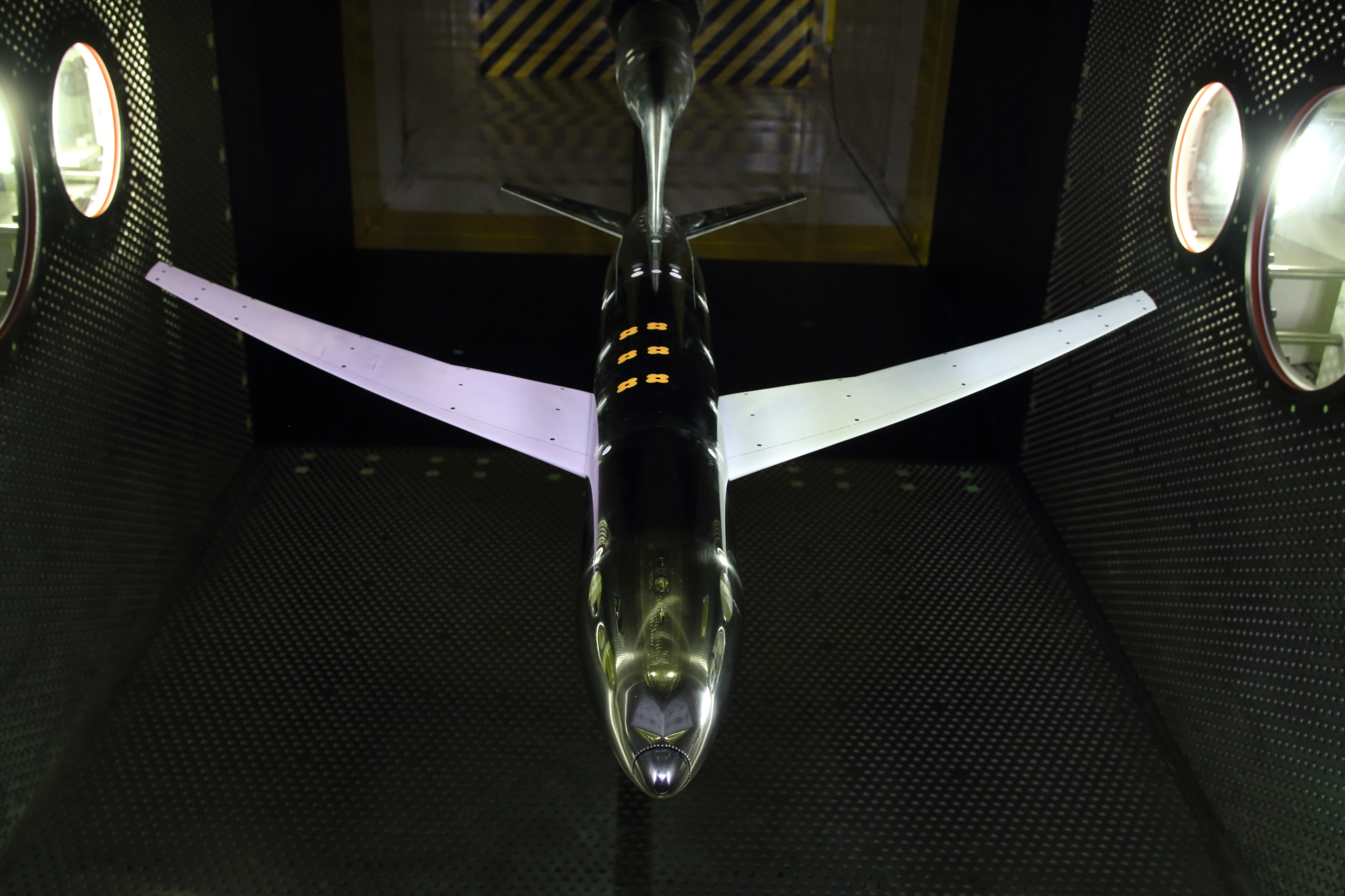}}
\caption{80\%-scale NASA Common Research Model.}\label{fig:crm_model}
\end{figure}

This study analyzes pressure coefficient ($C_P$) distribution on the main wing of the NASA common research model (NASA‑CRM) \cite{Vassberg2008}, obtained from the wind‑tunnel experiment conducted by Sugioka et al. \cite{Sugioka2021}.
An 80\%-scale model of the NASA-CRM that includes the main wing, fuselage, and horizontal tailplane shown in Fig. \ref{fig:crm_model} was used as a test article. 
The reference (mean aerodynamic) chord length and span length are $c_{\rm MAC} = 0.151~$m and $b = 1.269~$m, respectively. 
Further details of this model, please refer to \cite{Ueno2014}.
The experiments were conducted in JAXA $2\mathrm{m} \times 2\mathrm{m}$ transonic wind tunnel (JTWT1) at the JAXA Chofu Aerospace Center. The flow conditions were set to a Mach number of 0.85, with a Reynolds number of $2.27 \times 10^6$ based on the mean aerodynamic chord $c_{\rm MAC}$, and an angle of attack of $4.82^{\circ}$. Under these conditions, the flow exhibits both low-frequency shock-wave oscillations and buffet cell phenomena \cite{Sugioka2021}, which are the focus of this study.
Unsteady surface‑pressure fluctuations were measured with a fast‑response polymer/ceramic pressure‑sensitive paint (PC‑PSP) system \cite{Sugioka2018}.
The uncertainty of the obtained pressure signals was evaluated by comparison with unsteady pressure transducer data, and the noise floor was estimated to be approximately $10^{-4}$ in PSD [$(\Delta C_P)^2/St$] \cite{Sugioka2021}.
Further details on data processing, calibration, and uncertainty analysis are provided in \cite{Sugioka2021}.

The input data consist of pressure coefficient ($C_P$) distributions measured on the wing surface after adjusting the angle of attack of the model and confirming that the flow had settled.
For the present analysis, the $C_P$ distribution was mapped onto the suction side of the wing using linear interpolation, with a maximum grid spacing of approximately $1.7~\mathrm{mm}$ and about $6\times10^{4}$ grid points.
A total of $m = 1000$ snapshots were analyzed, with a time interval of $\Delta t = 0.38$ in dimensionless time units (normalized by the mean aerodynamic chord $c_{\rm MAC}$ and freestream velocity $U_\infty$) between consecutive snapshots.
This sampling yields a total measurement duration of 380 time units, capturing more than 20 cycles of the shock oscillations (assuming that the Strouhal number of the oscillation is $St = 0.06$).
The selected sampling rate provides approximately 4--13 samples per buffet cell oscillation period (with $St$ ranging from 0.2 to 0.6), ensuring sufficient temporal resolution to capture the dynamics of both the low‑frequency shock oscillations and the higher‑frequency buffet cell phenomena.

\begin{figure}
\centerline{\includegraphics[width=1.0\textwidth]{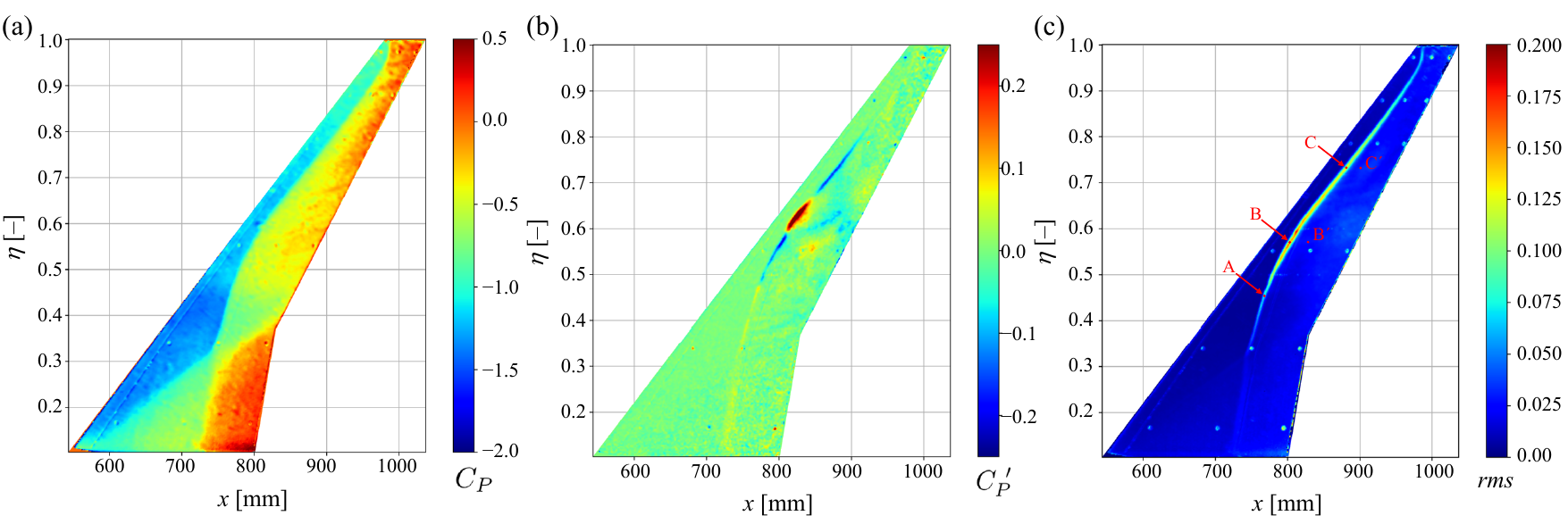}}
\caption{Pressure coefficient distributions on the suction side of the wing. (a) time-averaged; (b) instantaneous fluctuating component $C_P^\prime$; (c) root mean square of $C_P^\prime$. See supplementary movie 1.}\label{fig:inputCP}
\end{figure}
Fig.~\ref{fig:inputCP} presents the pressure coefficient distribution on the suction side of the wing. The time-averaged field (Fig.~\ref{fig:inputCP}(a)) shows insufficient $C_P$ recovery at the trailing edge in the midspan region. This suggests that the flow, which separates downstream of the shock-wave, remains detached at the trailing edge. Moving towards the outboard, $C_P$ begins to recover at the trailing edge, indicating that the separation is less severe than at the midspan location.

The fluctuating component (Fig.~\ref{fig:inputCP}(b)) shows large variations at the shock position from midspan to outboard. These spatial patterns arise from two phenomena: low-frequency shock-wave oscillation and buffet cells, which will be discussed in detail in subsequent sections. The root mean square (rms) distribution of the fluctuations is shown in Fig.~\ref{fig:inputCP}(c).
Large rms values are observed along the shock-wave, with a peak occurring in the midspan region near point B (at $\eta = 0.57$) where significant flow separation is present.
Here, $\eta = y / (b/2)$ is the spanwise location $y$ normalized by the half-span length $b/2$.
In this study, the time evolution data of the fluctuating field (Fig.~\ref{fig:inputCP}(b)) serves as the input for the VMD-NCS analysis.

\section{Results and discussion}
\subsection{Extracted structures} 
\begin{figure}
  \centerline{\includegraphics[width=1.0\textwidth]{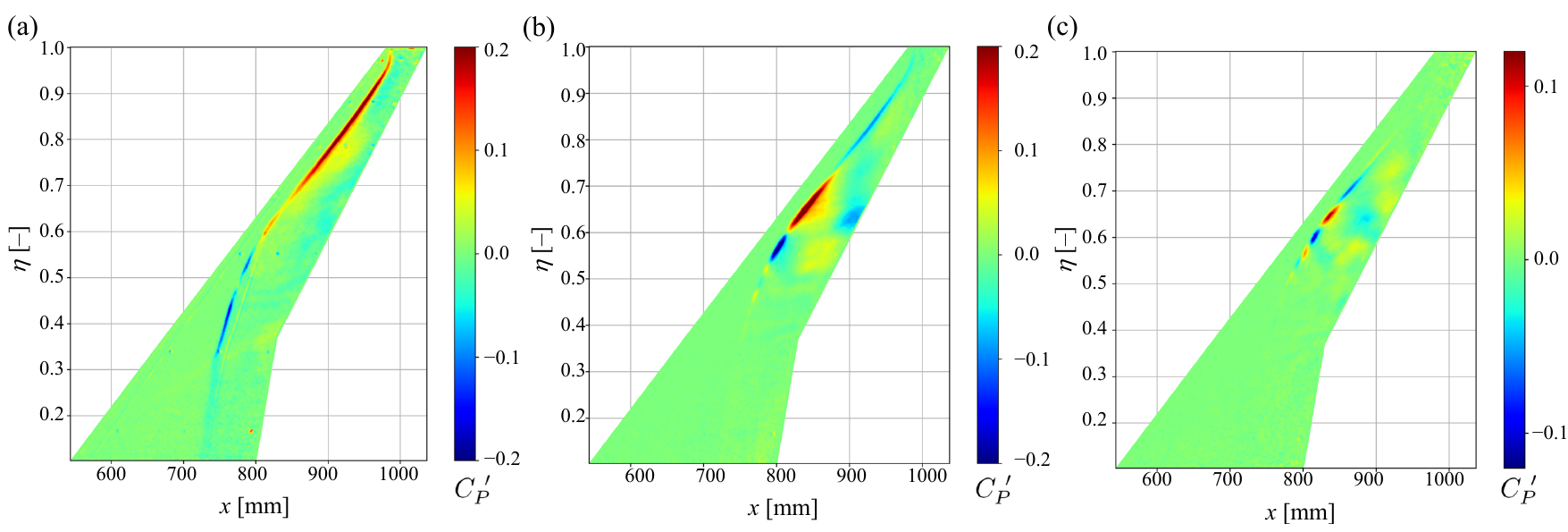}}
  \caption{Spatial distributions of the three intrinsic coherent structures (ICSs): (a) low-frequency shock oscillation, $ICS_1$ ($St_c=0.106$); (b) primary buffet cell, $ICS_2$ ($St_c=0.375$); (c) secondary buffet cell, $ICS_3$ ($St_c=0.795$). See supplementary movie 2.}
  \label{fig:ics_spatial}
\end{figure}

Fig.~\ref{fig:ics_spatial} shows the spatial distributions of three ICSs extracted using the VMD-NCS analysis. The center frequencies of these ICSs are $St_c = 0.106$, $0.375$, and $0.795$, referred to as $ICS_1$, $ICS_2$, and $ICS_3$, respectively. $ICS_1$ represents the low-frequency shock oscillation, while $ICS_2$ and $ICS_3$ capture the buffet cell structures. As the frequency increases from $ICS_2$ to $ICS_3$, the spatial wavelength of the buffet cells becomes shorter.

\begin{figure}
  \centerline{\includegraphics[width=1.0\textwidth]{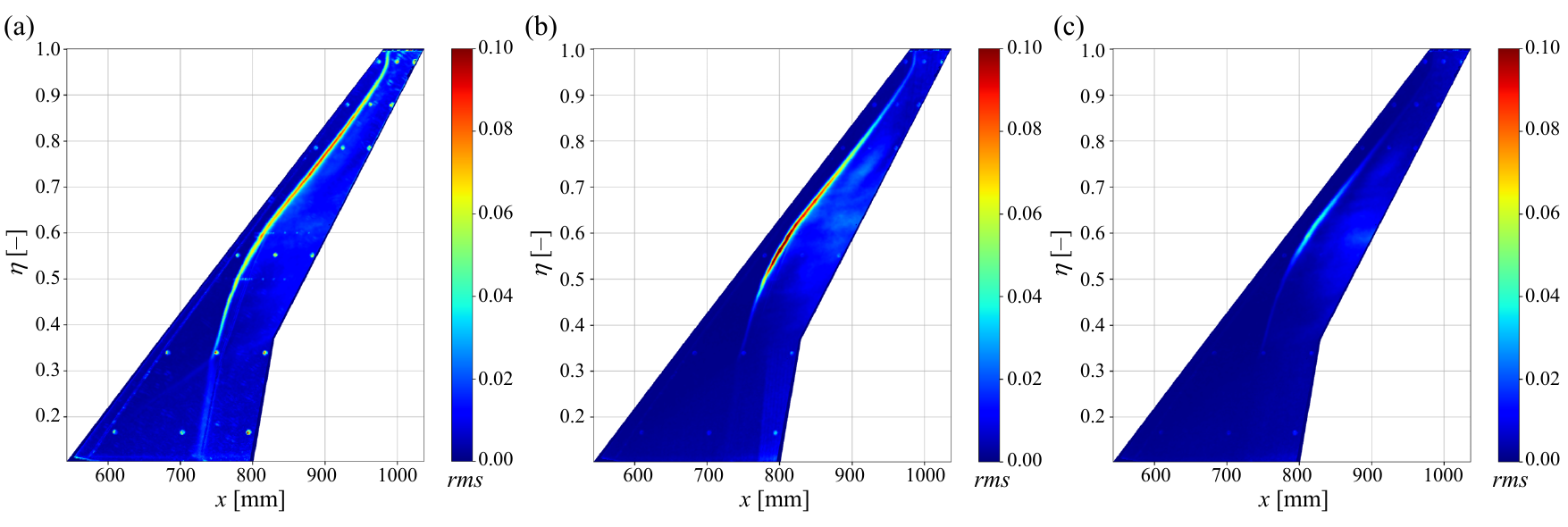}}
  \caption{Root mean square (rms) distributions of the intrinsic coherent structures (ICSs) on the wing surface: (a) low-frequency shock oscillation, $ICS_1$ ($St_c=0.106$); (b) primary buffet cell, $ICS_2$ ($St_c=0.375$); (c) secondary buffet cell, $ICS_3$ ($St_c=0.795$).}
  \label{fig:ics_rms}
\end{figure}

\begin{figure}
  \centerline{\includegraphics[width=1.0\textwidth]{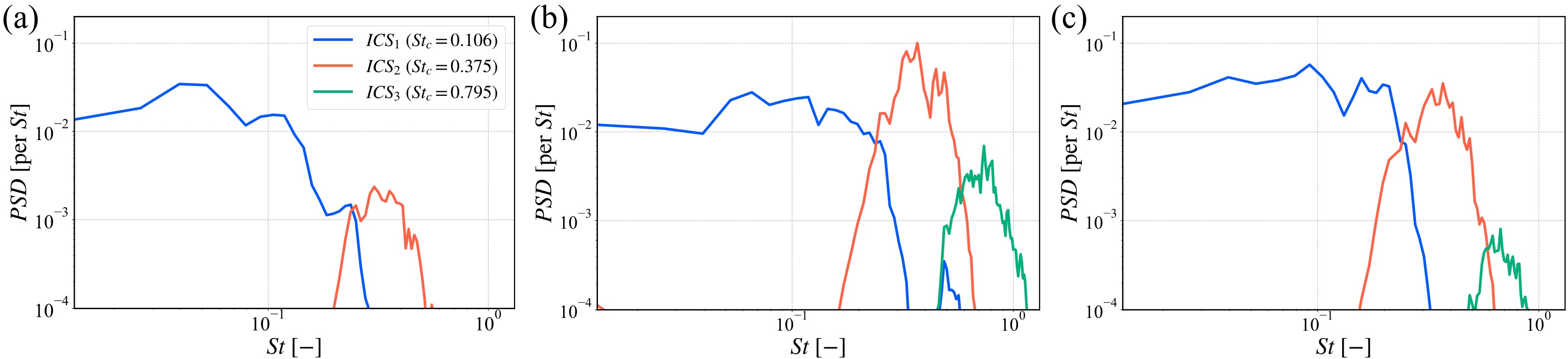}}
  \caption{Frequency distribution of pressure fluctuations ($C_P^\prime$) for each ICS at three spanwise locations: (a) $\eta = 0.45$; (b) $\eta = 0.57$; (c) $\eta = 0.73$. Measurement positions correspond to those marked as A, B, C in Fig.~\ref{fig:inputCP}(c).}
  \label{fig:ics_frequency}
\end{figure}

Figures~\ref{fig:ics_rms} and~\ref{fig:ics_frequency} illustrate the spanwise variation in the intensity and dominance of each ICS. Fig.~\ref{fig:ics_rms} shows the root mean square (rms) distributions of the ICSs across the wing surface, while Fig.~\ref{fig:ics_frequency} presents the power spectral densities (PSDs) of $C_P^\prime$ signals measured at three representative spanwise positions (A: $\eta = 0.45$, B: $\eta = 0.57$, C: $\eta = 0.73$). The PSDs were computed using Welch’s method with a Hann window; the signals were divided into nine segments with 50\% overlap to improve spectral estimation.
$ICS_2$ and $ICS_3$, which represent buffet cell structures, exhibit peak intensities in the midspan region (around  $\eta = 0.6$), as seen in both the rms and spectral data. This indicates that buffet cell fluctuations are most prominent in the midspan, where flow separation presents (Fig.~\ref{fig:inputCP}(a)).
In contrast, $ICS_1$, associated with low-frequency shock oscillation, is more dominant in the outboard region with its rms and PSD values.

These results also demonstrate the effectiveness of the VMD-NCS analysis in isolating frequency-specific fluctuation components while minimizing frequency mixing.
Unlike conventional POD, which often entangles multiple frequency components within a single mode \citep{Noack2016}, the VMD-NCS framework separates distinct spatiotemporal structures more clearly.
Relationship between POD and VMD-NCS analysis is discussed in Appendix B.

It should be noted that both $ICS_2$ and $ICS_3$ represent buffet cell structures, and their superposition can be interpreted as a single fluid phenomenon. In this study, $ICS_2$, which is dominant in terms of fluctuation intensity, was selected as the representative ICS for analyzing buffet cell behavior. However, we confirmed that the key findings and discussions remain consistent even when $ICS_2$ and $ICS_3$ are considered jointly as a single phenomenon (see Appendix C for details).

\subsection{Relationship between shock motion and buffet cells} \label{RelationSB}
\begin{figure}
  \centering
  \includegraphics[width=\textwidth]{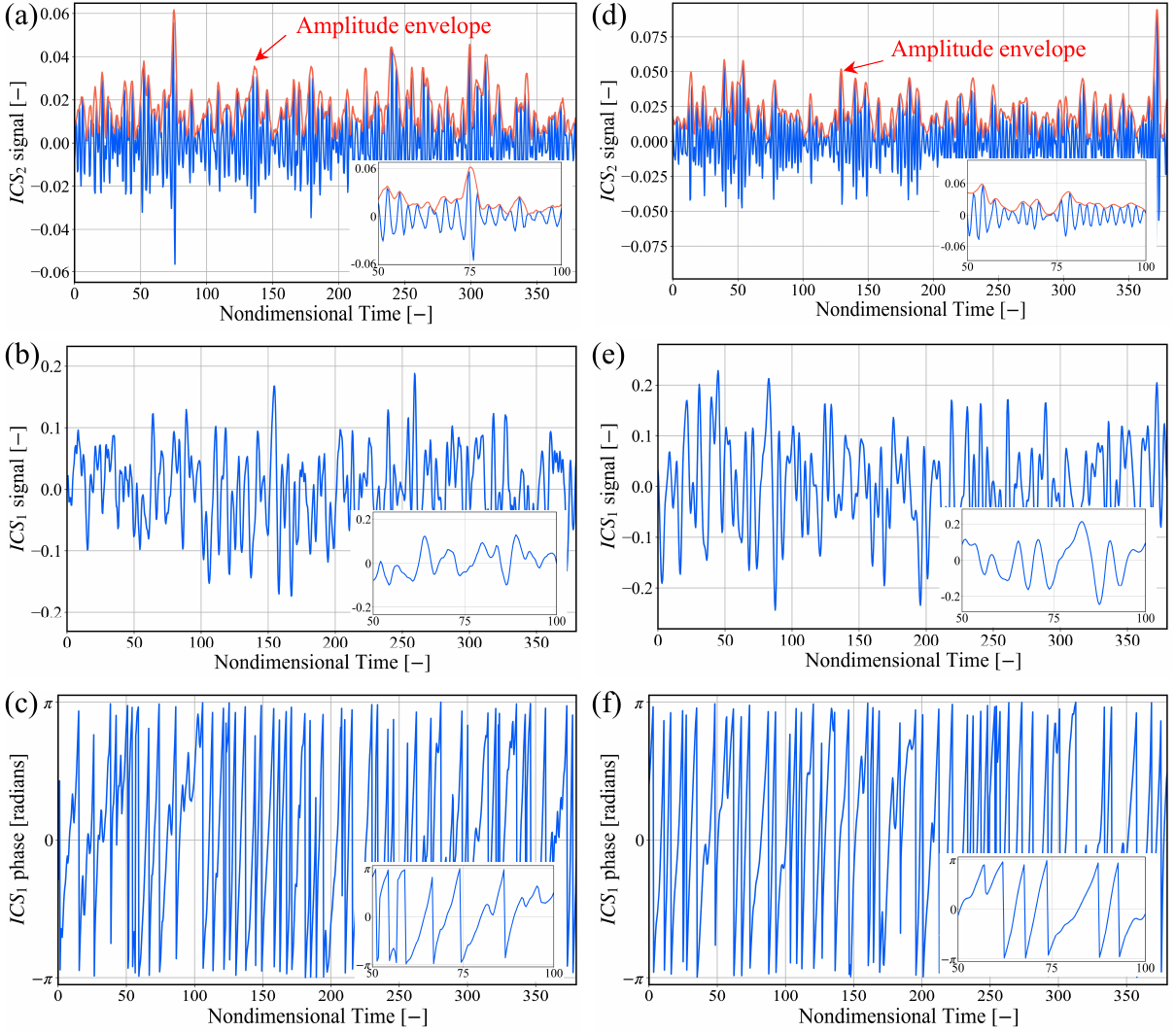}
  \caption{Time series of ICS signals at $\eta = 0.57$ (left column) and $\eta = 0.73$ 
    (right column): (a, d) $ICS_2$ signals with amplitude envelopes; (b, e) $ICS_1$ 
    signals; (c, f) $ICS_1$ phase. Each panel includes an inset showing a closeup view of the time interval 50--100.}
    \label{fig:ics_signals}
\end{figure}

\begin{figure}
  \centerline{\includegraphics[width=0.9\textwidth]{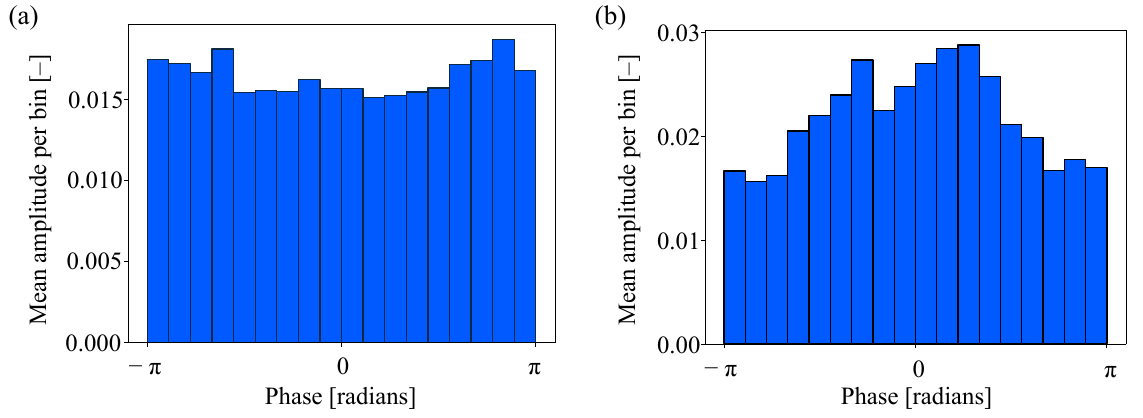}}
  \caption{Mean amplitude of buffet cell fluctuations ($ICS_2$) for each phase bin of low-frequency shock 
    oscillation ($ICS_1$): (a) $\eta = 0.57$; (b) $\eta = 0.73$. }
    \label{fig:histogram}
\end{figure}

Fig.~\ref{fig:ics_signals} shows pressure fluctuations of low-frequency shock oscillation ($ICS_1$) and buffet cells ($ICS_2$) at midspan ($\eta = 0.57$, (a)-(c)) and outboard ($\eta = 0.73$, (d)-(f)) positions.
The buffet cell fluctuations shown in figures~\ref{fig:ics_signals}(a) and (d) represent $ICS_2$ signals and their instantaneous amplitudes measured at locations $0.2c$ downstream of the shock-wave (points B$'$ and C$'$ in Fig.~\ref{fig:inputCP}(c)), where $c$ is the local chord length.
Figures~\ref{fig:ics_signals}(b) and (e) show the $ICS_1$ signals, while figures~\ref{fig:ics_signals}(c) and (f) show the instantaneous phase of $ICS_1$ at the shock foot (points B and C), corresponding to low-frequency shock oscillation.
The instantaneous amplitudes and phases were computed using Eq.~(\ref{Eq:amp_phase}) through Hilbert transform.
In the $ICS_1$ phase shown in figures~\ref{fig:ics_signals}(c) and (f), phases of 0 corresponds to the local maximum pressure rise caused by the shock oscillation, whereas phases of $\pm\pi$ corresponds to the local minimum (pressure decrease).
The figures show that both low-frequency shock-wave oscillation and buffet cell behaviors are nonstationary phenomena. The amplitude of buffet cell fluctuations ($ICS_2$) is modulated, but the relationship to the phase of the shock oscillation ($ICS_1$) is not clear from these time series alone.

The histograms in Fig.~\ref{fig:histogram} are created following the phase-amplitude coupling (PAC) visualization method developed by Tort et al. \cite{Tort2008}, based on the signals in Fig.~\ref{fig:ics_signals}. 
For each time instant, the instantaneous phase of $ICS_1$ and the instantaneous amplitude of $ICS_2$ are paired. 
The $ICS_1$ phase range $[-\pi, \pi]$ is then divided into 18 bins at $\pi/9$ intervals, and all amplitude values of $ICS_2$ corresponding to each phase bin are averaged to create the histogram. 
This approach directly visualizes how the amplitude of buffet cells varies with respect to the phase of shock oscillations, providing an intuitive representation of their interaction.
Note that, to validate that the discussion in this paper is not dependent on the specific choice of binning parameters, we performed an additional analysis by varying the number of phase bins from 12 to 24 and confirmed that the qualitative trends remained consistent across all tested bin configurations.

Fig.~\ref{fig:histogram}(a) shows that at the midspan section, the average amplitude of $ICS_2$ has little dependence on the phase of $ICS_1$. In other words, at this spanwise position where buffet cells are prominent (as shown in Fig.~\ref{fig:ics_frequency}(b)), the amplitude of buffet cell fluctuations ($ICS_2$) is not significantly affected by the phase of low-frequency shock-wave oscillation ($ICS_1$).
Conversely, Fig.~\ref{fig:histogram}(b) shows that at the outboard section where low-frequency shock oscillation ($ICS_1$) dominates over buffet cells ($ICS_2$), the average amplitude of $ICS_2$ depends significantly on the phase of $ICS_1$. Specifically, buffet cell fluctuations become stronger when the phase of low-frequency oscillation ($ICS_1$) is around 0, corresponding to the shock-wave being located upstream due to the low-frequency oscillation.

\begin{figure}
  \centerline{\includegraphics[width=1\textwidth]{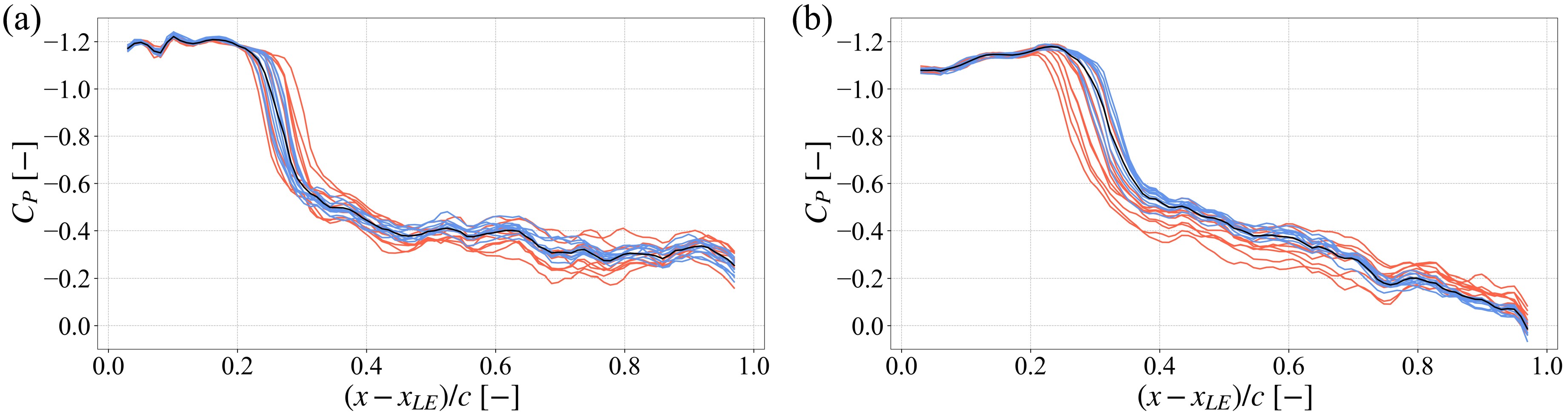}}
  \caption{Chordwise pressure distributions at moments of local maximum (red) and 
    minimum (blue) buffet cell amplitudes. Ten highest/lowest 
    peaks moments shown.
    The black line indicates the time-averaged pressure distribution.    
    (a) $\eta = 0.57$; (b) $\eta = 0.73$.}
    \label{fig:cpdist_bfamp}
\end{figure}

Fig.~\ref{fig:cpdist_bfamp} presents the chordwise pressure distributions at moments when the amplitude of buffet cell fluctuations ($ICS_2$) reaches local maxima (red lines) and minima (blue lines), corresponding to the peak and trough points in figures~\ref{fig:ics_signals}(a) and (d). The black lines indicate the time-averaged pressure distributions for reference.
The chordwise coordinate is normalized by the leading edge position $x_{\rm LE}$ and local chord length $c$.
At the midspan location (Fig.~\ref{fig:cpdist_bfamp}(a)),  the shock-wave position, identified by a steep pressure rise, fluctuates regardless of whether the buffet cell amplitude is high or low, appearing both upstream and downstream of the mean location.
This suggests a lack of clear correlation between the shock-wave position and the amplitude of buffet cell fluctuations at this spanwise location.
Additionally, the variation in shock position tends to increase when the buffet cell amplitude is high.
As shown in figures~\ref{fig:ics_rms} and~\ref{fig:ics_frequency}, this region is dominated by buffet cell fluctuations.
These results suggest that the shock displacement in this region is primarily driven by buffet cells rather than by low-frequency shock oscillations.
Moreover, the consistently flat pressure distribution downstream of the shock suggests persistent boundary layer separation in this region.

In contrast, the outboard location (Fig.~\ref{fig:cpdist_bfamp}(b)) exhibits a clear correlation between the shock position and the buffet cell amplitude.
When the buffet cell amplitude is high (red lines), the shock is located upstream, while it stays downstream when the amplitude is low (blue lines).
This phase-aligned movement indicates a strong coupling between shock-wave position and buffet cell strength.
Furthermore, when the shock is downstream, the pressure recovers toward the trailing edge, indicating attached flow.
When the shock is upstream, the pressure gradient behind it becomes shallower, implying flow separation.
These observations suggest that, in the outboard region, the boundary layer separation state varies with shock position and modulates buffet cell amplitude.

Based on these findings, the phase of low-frequency shock-wave oscillations does not directly control buffet cell fluctuations, as evidenced by the lack of phase dependency at the midspan location where buffet cells are most prominent. Instead, buffet cells are more closely correlated with the boundary layer separation state.
Strong buffet cell fluctuations correspond to separated flow conditions, while their amplitude diminishes under attached flow.
Moreover, in regions where buffet cells dominate, the chordwise shock position is primarily influenced by buffet cell fluctuations rather than low-frequency shock oscillations.
Taken together, the results indicate that while shock-waves are a key factor in generating the adverse pressure gradient required for boundary layer separation, their influence on buffet cells is mediated through the resulting separation behavior.
This observed dependence of buffet cell behaviors on boundary layer separation characteristics aligns with the computational findings of Plante et al. \cite{Plante2021}, which suggested through global stability analysis that buffet cells and stall cells emerge from similar flow mechanisms related to the boundary layer separation. Additionally, the recent high-fidelity simulations by Lusher et al. \cite{Lusher2024} reported that buffet cells become prominent only when extensive mean flow separation is present, which is consistent with our observation that strong buffet cell fluctuations occur when the pressure gradient behind the shock-wave becomes flat, indicating separation.

\section{Conclusions}
This study investigated the relationship between low-frequency shock-wave oscillations and buffet cells on a swept wing by applying VMD-NCS analysis to unsteady pressure-sensitive paint (PSP) data obtained from wind tunnel tests. The analysis successfully extracted coherent structures corresponding to low-frequency shock motion and higher-frequency buffet cells, enabling detailed examination of their spatial and temporal characteristics.

The results revealed that the behavior of buffet cells is associated with the boundary layer separation state rather than the phase of shock-wave oscillations.
In regions where the boundary layer remains persistently separated around the midspan, buffet cell fluctuations dominate independently of shock motion. In contrast, in the outboard region where the separation state changes in response to shock movement, the amplitude of buffet cell fluctuations varies in phase with shock-wave displacement. These findings indicate that while shock-waves generate the adverse pressure gradient necessary for separation, the manifestation of buffet cell fluctuations is dictated by the separation dynamics of the boundary layer.

The experimental observations in this study align with previous computational findings, including those by Plante et al. \cite{Plante2021}, which suggested that buffet cells and stall cells emerge from similar flow mechanisms related to boundary layer separation, and Lusher et al. \cite{Lusher2024}, which reported that buffet cells become prominent only when extensive mean flow separation is present. 
While the present results are based on a single experimental dataset at specific conditions ($M = 0.85$, $Re = 2.27 \times 10^6$, $\alpha = 4.82^{\circ}$), the consistency with these computational studies on different configurations suggests that the relationship between buffet cells and separation characteristics may be a general feature of buffet cell phenomena.
Future experimental studies across various flow conditions and geometries would be valuable to confirm this generality. Furthermore, while this study demonstrates strong correlations between buffet cells and boundary layer separation state, establishing definitive causal mechanisms would require additional investigations.

Additionally, these results underscore the capability of VMD-NCS to isolate nonstationary structures with minimal frequency mixing, proving its effectiveness in analyzing complex unsteady flow phenomena.

\section*{Acknowledgment}
This work was supported in part by JST PRESTO Grant Number JPMJPR23O2, Japan.

\section*{Appendix A. Effect of $K$ and $\alpha$ on frequency separation in VMD-NCS analysis}
\begin{figure}
  \centering
  \includegraphics[width=\textwidth]{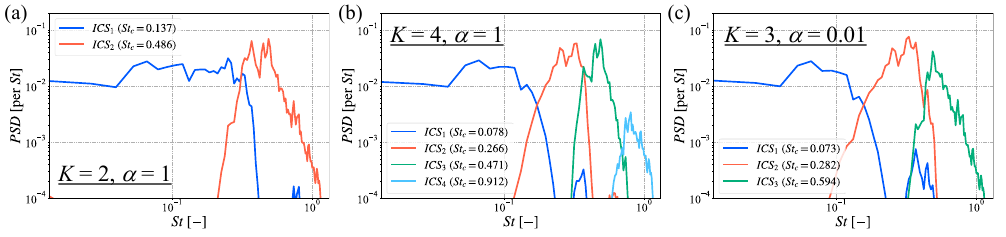}
  \caption{Effect of $K$ and $\alpha$ on ICSs frequency distribution. Power spectral densities of pressure fluctuations ($C_P^\prime$) for each ICS at point B (marked in Fig.~\ref{fig:inputCP}(c)) with different $K$ and $\alpha$.}
  \label{fig:eop}
\end{figure}

The parameters $K$ (number of ICSs) and $\alpha$ (penalty factor) in VMD-NCS analysis influence the frequency distribution of the extracted ICSs. To ensure appropriate parameter selection, we investigated the frequency distributions of ICSs under different combinations of $K$ and $\alpha$ values.

Figure~\ref{fig:eop} shows the power spectral densities of ICSs at the midspan location where buffet cells are prominent.
For the $K=2$ case (Fig.~\ref{fig:eop}(a)), the center frequency of $ICS_1$ is closer to the buffet cell frequency range (bump-shaped peak around $St=0.2$--0.6) compared to the $K=3$ case (Fig.~\ref{fig:ics_frequency}(b)), which degrades the separation of shock oscillations and buffet cells.
For the $K=4$ case (Fig.~\ref{fig:eop}(b)), the increased number of ICSs leads to excessive fragmentation, where the single bump-shaped peak corresponding to buffet cells is unnecessarily divided into three separate ICSs.

When $\alpha$ is reduced (Fig.~\ref{fig:eop}(c)), the bump-shaped peak is divided near its peak frequency ($St \approx 0.4$), and each ICS exhibits broader bandwidth compared to the $\alpha=1$ case (Fig.~\ref{fig:ics_frequency}(b)). For larger values of $\alpha$ (e.g., $\alpha = 100$), the MVMD iterations failed to converge.

It should be noted that the reconstruction error remained sufficiently small (approximately 0.3\% in relative error, Eq.~(\ref{eq:reconst_error})) for all converged cases, indicating that reconstruction accuracy alone cannot differentiate between parameter choices.

Based on these observations, we selected $K=3$ and $\alpha=1$ for this study, as this combination provided the clearest separation between buffet cells and low-frequency shock oscillations. However, even with this optimal setting, the bump-shaped peak is represented by two ICSs, which is further discussed in Appendix~C.

\section*{Appendix B. Relationship between POD and VMD-NCS analysis}
\begin{figure}
  \centering
  \includegraphics[width=\textwidth]{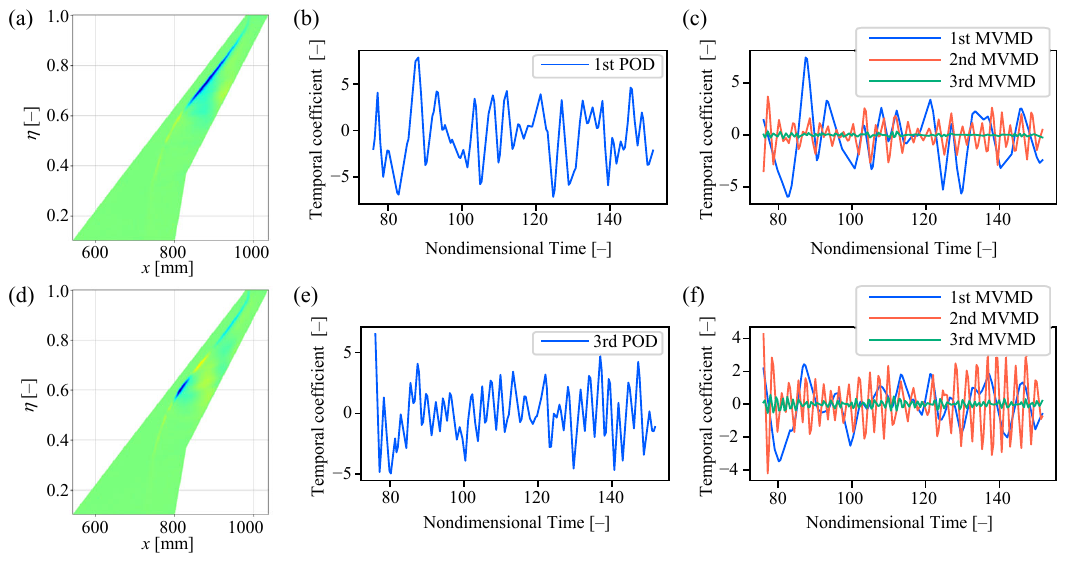}
  \caption{(a), (d) 1st and 3rd POD mode shapes; (b), (e) Temporal coefficients of the 1st and 3rd POD modes; (c), (f) Decomposed temporal coefficients obtained by applying MVMD to the 1st and 3rd POD temporal coefficients.}
    \label{fig:pod_comparison}
\end{figure}
In this section, we explain the relationship between POD and VMD-NCS analysis. Fig.~\ref{fig:pod_comparison} illustrates the representative POD mode shapes (1st and 3rd), their POD temporal coefficients $\tilde q_i(t)$, and the corresponding MVMD modes $\tilde u_{k,i}(t)$ (Eq.~\ref{eq:MVMDofPOD}).
Figures \ref{fig:pod_comparison}(a) and (d) appear to suggest that POD extracts patterns that capture the characteristics of low-frequency shock oscillation and buffet cells. However, their temporal coefficients (figures \ref{fig:pod_comparison}(b) and (e)) exhibit complex behavior due to the mixing of multiple frequency components.
In VMD-NCS analysis, by applying MVMD to the POD temporal coefficients, we further separate the fluctuations into distinct frequency bands (figures \ref{fig:pod_comparison}(c) and (f)). 
This decomposition of the temporal coefficients using MVMD enables the representation of spatiotemporal structures based on POD mode shapes and MVMD modes (Eq.~\ref{eq:ICS}), allowing patterns in each frequency band to be clearly extracted (as shown in Fig.~\ref{fig:ics_frequency}).

\section*{Appendix C. Combination of ICSs for buffet cell extraction}
\begin{figure}
  \centerline{\includegraphics[width=1\textwidth]{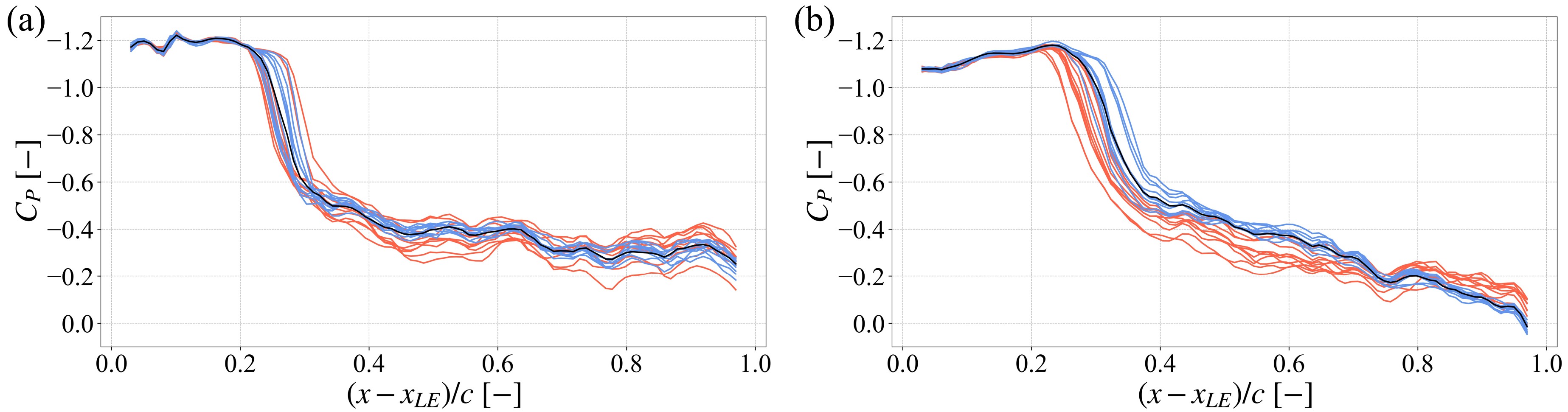}}
  \caption{Same as Fig.~\ref{fig:cpdist_bfamp} but buffet cell signals are extracted as the combined ICSs ($ICS_2+ICS_3$).}
    \label{fig:cpdist_ICS23}
\end{figure}
In Section~\ref{RelationSB}, $ICS_2$ was used to represent the buffet cell pattern to analyze the relationship between low-frequency shock oscillations and buffet cells. However, both $ICS_2$ and $ICS_3$ capture variations in buffet cell behavior, and it is reasonable to consider that the buffet cell fluctuation can be represented as a combination of these two ICSs.

Therefore, in this study, we also performed an analysis using the combined buffet cell signal, $ICS_2 + ICS_3$. Fig.~\ref{fig:cpdist_ICS23} shows the results of this analysis, which is identical to the analysis performed in Fig.~\ref{fig:cpdist_bfamp} but with $ICS_2 + ICS_3$ as the buffet cell signal.
The distribution in Fig.~\ref{fig:cpdist_ICS23} is nearly identical to that in Fig.~\ref{fig:cpdist_bfamp}, indicating that using the combined signal to represent the buffet cell does not affect the conclusions drawn from this study.

\bibliographystyle{elsarticle-num}

\bibliography{ref}

\end{document}